\documentclass[useAMS,usenatbib]{mn2e}
\usepackage{graphicx}
\input psfig.sty

\title[The progenitor of type Ia supernovae with different metallicities]
{The single-degenerate channel for the progenitor of type Ia supernovae with different metallicities}
\author[Meng, Chen and Han]
       {X. Meng$^{\rm 1,2}$ \thanks{E-mail:
conson859@msn.com}, X. Chen$^{\rm 1}$ and Z. Han $^{\rm 1}$  \\
        $^1$ National Astronomical Observatories/Yunnan
Observatory, the Chinese Academy of Sciences, Kunming, 650011,
China, \\conson859@msn.com\\
$^2$ Department of Physics and Chemistry, Henan Polytechnic
University, Jiaozuo, 454000, China}

\begin{document}

\date{Accepted. Received}

\pagerange{\pageref{firstpage}--\pageref{lastpage}} \pubyear{2007}

\maketitle

\label{firstpage}

\begin{abstract}
The single-degenerate channel for the progenitors of type Ia
supernovae (SNe Ia) are currently accepted, in which a
carbon-oxygen white dwarf (CO WD) accretes hydrogen-rich material
from its companion, increases its mass to the Chandrasekhar mass
limit, and then explodes as a SN Ia. Incorporating the
prescription of \citet{HAC99a} for the accretion efficiency into
Eggleton's stellar evolution code and assuming that the
prescription is valid for \emph{all} metallicities, we performed
binary stellar evolution calculations for more than 25,000 close
WD binaries with metallicities $Z=0.06, 0.05, 0.04, 0.03, 0.02,
0.01, 0.004, 0.001, 0.0003$ and $0.0001$. For our calculations,
the companions are assumed to be unevolved or slightly evolved
stars (WD + MS). As a result, the initial parameter spaces for SNe
Ia at various $Z$ are presented in orbital period-secondary mass
($\log P_{\rm i}, M_{\rm 2}^{\rm i}$) planes. Our study shows that
both the initial mass of the secondary and the initial orbital
period increase with metallicity. Thus, the minimum mass of the CO
WD for SNe Ia decreases with metallicity $Z$. The difference of
the minimum mass may be as large as 0.24 $M_{\odot}$ for different
Z.

Adopting the results above, we studied the birth rate of SNe Ia
for various $Z$ via a binary population synthesis approach. If a
single starburst is assumed, SNe Ia occur systemically earlier and
the peak value of the birth rate is larger for a high $Z$. The
Galactic birth rate from the WD + MS channel is lower than (but
comparable to) that inferred from observations. Our study
indicates that supernovae like SN2002ic would not occur in
extremely low-metallicity environments, if the delayed
dynamical-instability model in \citet{HAN06} is appropriate.

\end{abstract}

\begin{keywords}binaries:close-stars:evolution-supernovae:general-white
dwarfs
\end{keywords}

\section{Introduction}\label{sect:1}
Type Ia supernovae (SNe Ia) play an important role in
astrophysics, especially in cosmology. They appear to be good
cosmological distance indicators and are successfully applied to
determine cosmological parameters (e.g. $\Omega$ and $\Lambda$;
\citealt{RIE98}; \citealt{PER99}). There is a linear relation
between the absolute magnitude of SNe Ia at maximum light and the
magnitude drop of B light curve during the first 15 days following
maximum. This relation is now known as \textbf{the} Phillips
relation (\citealt{PHI93}) and \textbf{is} adopted when SNe Ia
\textbf{are used} as distance indicators. In this case, the
Phillips relation is assumed to be valid at high redshift,
although it comes from low-redshift samples. Obviously, this
assumption is precarious since the exact nature of SNe Ia is still
unclear (see the reviews by \citealt{HN00}; \citealt{LEI00}). If
the properties of SNe Ia evolve with redshift, the results for
cosmology might be different. Thus, it is necessary to study the
properties of SNe Ia at high redshift. Since metallicity may
represent redshift to some extent (e.g. metallicity decreases with
redshifts), it is a good method to study the properties of SNe Ia
at various redshift by finding correlations between their
properties and metallicity.

It is generally agreed that SNe Ia originate from the
thermonuclear runaway of a carbon-oxygen white dwarf (CO WD) in a
binary system. The CO WD accretes material from its companion,
increases mass to its maximum stable mass, and then explodes as a
thermonuclear runaway. Almost half of the WD mass is converted
into radioactive nickel-56 in the explosion (\citealt{BRA04}), and
the amount of nickel-56 determines the maximum luminosity of SNe
Ia (\citealt{ARN82}). Some numerical and synthetical results
showed that metallicity has an effect on the final amount of
nickel-56, and thus the maximum luminosity (\citealt{TIM03};
\citealt{TRA05}; \citealt{POD06}). There is also some other
evidence of the correlation between the properties of SNe Ia and
metallicity from observations, e.g. \citealt{BB93};
\citealt{HAM96}; \citealt{WAN97}; \citealt{CAP97};
\citealt{SHA02}.

Although the fact that SNe Ia originate from the thermonuclear
disruption of mass accreting white dwarfs is widely accepted, the
precise nature of the progenitor systems remains unclear.
According to the nature of the companions of the mass accreting
white dwarfs, two competing scenarios have been proposed, i.e the
double-degenerate channel (DD, \citealt{IT84}; \citealt{WI87}) and
the single degenerate channel (SD, \citealt{WI73};
\citealt{NTY84}). In the DD channel, two CO WDs with a total mass
larger than the Chandrasekhar mass limit may coalesce, and then
explode as a SN Ia. Although the channel is theoretically less
favored, e.g. double WD mergers may lead to accretion-induced
collapses rather than to SNe Ia (\citealt{HN00}), it is premature
to exclude the channel at present since there exists evidence that
the channel may contribute a few SNe Ia (\citealt{HOW06};
\citealt{BRA06}; \citealt{QUI07}). The single-degenerate
Chandrasekhar model is the most widely accepted one at present
(\citealt{WI73}; \citealt{NTY84}). In this model, the maximum
stable mass of a CO WD is $\sim 1.378 M_{\odot}$ (close to the
Chandrasekhar mass, \citealt{NTY84}), and the companion is
probably a main sequence star or a slightly evolved star (WD+MS),
or a red-giant star (WD+RG) (\citealt{YUN95}; \citealt{LI97};
\citealt{HAC99a, HAC99b}; \citealt{NOM99}; \citealt{LAN00};
\citealt{HAN04}). The SD model is supported by some observations.
For example, variable \textbf{circumstellar} absorption lines were
observed in the spectra of SN Ia 2006X (\citealt{PAT07}), which
indicates the SD nature of its precursor. \citet{PAT07} suggested
that the progenitor of SN 2006X is a WD + RG system based on the
expansion velocity of the \textbf{circumstellar} material, while
\citet{HKN08} argued a WD + MS nature for this SN Ia. Recently,
\citet{VOSS08} suggested that SN 2007on is also possibly from a WD
+ MS channel. Moreover, several WD + MS systems are possible
progenitors of SN Ia. For example, supersoft X-ray sources (SSSs)
were suggested as good candidates for the progenitors of SNe Ia
(\citealt{HK03a, HK03b}). Some of the SSSs are WD + MS systems and
some are WD + RG systems (\citealt{DIK03}). A direct way to
confirm the progenitor model is to search for the companion stars
of SNe Ia in their remnants. The discovery of the potential
companion of Tycho's supernova may have verified the reliability
of the WD + MS model (\citealt{RUI04}; \citealt{IHA07}). In this
paper, we only focus on the WD + MS channel, which is a very
important channel for producing SNe Ia in the Galaxy
(\citealt{HAN04}).

There are two problems for the WD + MS channel when confronted
with observations, which are the absence of H in the vast majority
of SNe Ia spectra and the difficulty of producing SNe Ia with long
delay times (older then 10 Gyr inferred from SNe Ia in elliptical
galaxies in the local universe, \citealt{MAN05}) or extremely
short delay times (shorter than 0.1 Gyr, \citealt{MAN06}). The
absence of H in the spectra likely implies that the density of
circumstellar material (CSM) around SNe Ia is very low or the
amount of the hydrogen-rich material stripped from the companion
is very small (\citealt{MAT05}; \citealt{LEO07}). Neither of the
two problems can be overcome easily in the WD + MS channel. We
know from observations that some SNe Ia have delay times longer
than 10 Gyr or shorter than 0.1 Gyr (\citealt{SB05};
\citealt{MAN06}; \citealt{TOTANI08}), but the WD + MS channel can
only account for SNe Ia with delay times shorter than 2 Gyr while
longer than 0.1 Gyr (\citealt{HAN04}). It is likely that SNe Ia
with long or short delay times come from other channels than the
WD + MS channel, e.g. SNe Ia with long delay times come from WD
+RG channel, while those with short delay times come from a WD +
He star channel (Wang, Meng, Chen \& Han 2008, in preparation).

{Many works have concentrated on the WD+MS channel. \citet{HAC99a,
HAC99b, HKN08} and \citet{NOM99, NOM03} have studied the WD+MS
channel by a simple analytical method for treating binary
interactions. Such analytic prescriptions, as pointed out by
\citet{LAN00}, can not describe some mass-transfer phases,
especially those occurring on a thermal time-scale. \citet{LI97}
studied this channel from detailed binary evolution calculation,
while two WD masses, 1.0 and 1.2 $M_{\odot}$, are considered.
\citet{LAN00} investigated the channel for metallicities $Z=0.001$
and 0.02, but only for mass transfer during core hydrogen burning
phase (case A). \citet{HAN04} carried out a detailed study of the
channel including case A and early case B (mass transfer occurs at
Hertzprung gap (HG)) for $Z=0.02$. Considering that not all SNe Ia
are found in solar metallicity environments, we pay attention to
the correlation between the properties of SNe Ia and metallicities
here.

In this paper, we study the WD + MS channel comprehensively and
systematically at various Z, showing the parameter spaces for the
progenitors of SNe Ia and the distributions of the initial
parameters for the progenitors of SNe Ia. The results can be
\textbf{effectively} used in binary population synthesis, or be
helpful to search for the potential progenitor systems of SNe Ia.
In section \ref{sect:2}, we simply describe the numerical code for
binary evolution calculations and the grid of binary models we
have calculated. The evolutionary results are shown in section
\ref{sect:3}. We describe the binary population synthesis (BPS)
method in section \ref{sect:4} and present the BPS results in
section \ref{sect:5}. In section \ref{sect:6}, we briefly discuss
our results, and finally we summarize the main results in section
\ref{sect:7}.

\section[]{BINARY EVOLUTION CALCULATION}\label{sect:2}
We use the stellar evolution code of \citet{EGG71, EGG72, EGG73} to
calculate the binary evolutions of WD+MS systems. The code has been
updated with the latest input physics over the last three decades
(\citealt{HAN94}; \citealt{POL95, POL98}). Roche lobe overflow
(RLOF) is treated within the code described by \citet{HAN00}. We set
the ratio of mixing length to local pressure scale height,
$\alpha=l/H_{\rm p}$, to 2.0, and set the convective overshooting
parameter, $\delta_{\rm OV}$, to 0.12 (\citealt{POL97};
\citealt{SCH97}), which roughly corresponds to an overshooting
length of $0.25 H_{\rm P}$. Ten metallicities are adopted here (i.e.
$Z=$0.0001, 0.0003, 0.001, 0.004, 0.01, 0.02, 0.03, 0.04, 0.05 and
0.06). The opacity tables for these metallicties are compiled by
\citet{CHE07} from \citet{IR96} and \cite{AF94}. For each $Z$, the
initial hydrogen mass fraction is obtained by

 \begin{equation}
 X=0.76-3.0Z,
  \end{equation}
(Pols et al. 1998). Based on the correlation among $X$, $Y$ and
$Z$ used here, Pols et al. (1998) well reproduced the
color-magnitude diagram (CMD) of some clusters.

Instead of solving stellar structure equations of a WD, we adopt
the prescription of \citet{HAC99a} on WDs accreting hydrogen-rich
material from their companions. The following is a brief
introduction of this prescription. In the WD + MS channel, the
companion fills its Roche lobe at MS or during HG, and transfers
material onto the WD. If the mass-transfer rate, $|\dot{M}_{\rm
2}|$, exceeds a critical value, $\dot{M}_{\rm cr}$, we assume that
the accreted hydrogen steadily burns on the surface of WD and that
the hydrogen-rich material is converted into helium at the rate of
$\dot{M}_{\rm cr}$. The unprocessed matter is assumed to be lost
from the system as an optically thick wind at a rate of
$\dot{M}_{\rm wind}=|\dot{M}_{\rm 2}|-\dot{M}_{\rm cr}$
(\citealt{HAC96}). Based on the opacity from \citet{IR96}, the
optically thick wind is very sensitive to Fe abundance, and it is
likely that the wind does not work when $Z$ is lower than a
certain value (i.e. $Z< 0.002$, \citealt{KOB98}). Thus, there
should be an obvious low-metallicity threshold for SNe Ia in
comparison with SN II. However, this metallicity threshold has not
been found (\citealt{PRI07a}). Considering the uncertainties in
the opacities, we therefore assume rather arbitrarily that the
optically thick wind is still valid for low metallicities.

\begin{figure}
\centerline{\psfig{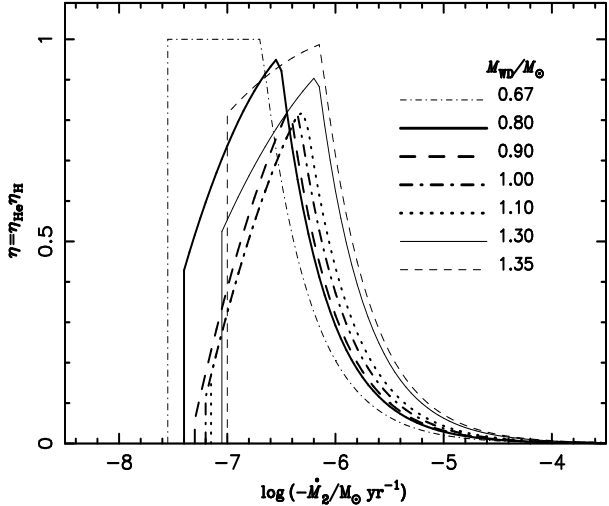}}
\caption{The total accumulation efficiency, $\eta=\eta_{\rm
He}\eta_{\rm H}$, as a function of mass-transfer rate,
$|\dot{M}_{\rm 2}|$. The hydrogen mass fraction, $X=0.7$, is
adopted for the critical accretion rate, $\dot{M}_{\rm cr}$.}
\label{eta}
\end{figure}

The critical accretion rate is

 \begin{equation}
 \dot{M}_{\rm cr}=5.3\times 10^{\rm -7}\frac{(1.7-X)}{X}(M_{\rm
 WD}-0.4),
  \end{equation}
where $X$ is hydrogen mass fraction and $M_{\rm WD}$ is the mass
of the accreting WD (mass is in $M_{\odot}$ and mass-accretion
rate is in $M_{\odot}/{\rm yr}$, \citealt{HAC99a}). We have not
included the effect of metallicities on equation (2) since the
effect is very small and can be neglected (\citealt{MEN06}).

The following assumptions are adopted when $|\dot{M}_{\rm 2}|$ is
smaller than $\dot{M}_{\rm cr}$. (1) When $|\dot{M}_{\rm 2}|$  is
higher than $\frac{1}{2}\dot{M}_{\rm cr}$, the hydrogen-shell
burning is steady and no mass is lost from the system. (2) When
$|\dot{M}_{\rm 2}|$ is lower than $\frac{1}{2}\dot{M}_{\rm cr}$
but higher than $\frac{1}{8}\dot{M}_{\rm cr}$, a very weak shell
flash is triggered but no mass is lost from the system. (3) When
$|\dot{M}_{\rm 2}|$ is lower than $\frac{1}{8}\dot{M}_{\rm cr}$,
the hydrogen-shell flash is so strong that no material is
accumulated on the surface of the CO WD. We define the growth rate
of the mass of the helium layer under the hydrogen-burning shell
as
 \begin{equation}
 \dot{M}_{\rm He}=\eta _{\rm H}|\dot{M}_{\rm 2}|,
  \end{equation}
where $\eta _{\rm H}$ is the mass accumulation efficiency for
hydrogen burning. According to the assumptions above, the values
of $\eta _{\rm H}$ are:

 \begin{equation}
\eta _{\rm H}=\left\{
 \begin{array}{ll}
 \dot{M}_{\rm cr}/|\dot{M}_{\rm 2}|, & |\dot{M}_{\rm 2}|> \dot{M}_{\rm
 cr},\\
 1, & \dot{M}_{\rm cr}\geq |\dot{M}_{\rm 2}|\geq\frac{1}{8}\dot{M}_{\rm
 cr},\\
 0, & |\dot{M}_{\rm 2}|< \frac{1}{8}\dot{M}_{\rm cr}.
\end{array}\right.
\end{equation}

Helium is ignited when a certain amount of helium is accumulated.
If a He-flash occurs, some of the helium is blown off from the
surface of the CO WD. Then, the mass growth rate of the CO WD,
$\dot{M}_{\rm WD}$, is
 \begin{equation}
 \dot{M}_{\rm WD}=\eta_{\rm He}\dot{M}_{\rm He}=\eta_{\rm He}\eta_{\rm
 H}|\dot{M}_{\rm 2}|,
  \end{equation}
where $\eta_{\rm He}$ is the mass accumulation efficiency for
helium-shell flashes, and its value is taken from \citet{KH2004}:

 \begin{equation}
  \eta_{\rm He}=1,
  \end{equation}
for $M_{\rm WD}< 0.8M_{\odot}$.

 \begin{equation}
\eta _{\rm He}=\left\{
 \begin{array}{lc}
-0.35(\log\dot{M}_{\rm He}+6.1)^{\rm 2}+1.02,\\
  \hspace{2.3cm}\log\dot{M}_{\rm He}<-6.34,\\
 1, \hspace{2.0cm}\log\dot{M}_{\rm He}\geq-6.34,\\
\end{array}\right.
\end{equation}
for $0.8 M_{\odot}\leq M_{\rm WD}< 0.9M_{\odot}$.

 \begin{equation}
\eta _{\rm He}=\left\{
 \begin{array}{lc}
-0.35(\log\dot{M}_{\rm He}+5.6)^{\rm 2}+1.07,\\
  \hspace{2.3cm}\log\dot{M}_{\rm He}<-6.05,\\
 1, \hspace{2.0cm}\log\dot{M}_{\rm He}\geq-6.05,\\
\end{array}\right.
\end{equation}
for $0.9 M_{\odot}\leq M_{\rm WD}< 1.0M_{\odot}$.
 \begin{equation}
\eta _{\rm He}=\left\{
 \begin{array}{lc}
-0.35(\log\dot{M}_{\rm He}+5.6)^{\rm 2}+1.01,\\
  \hspace{2.3cm}\log\dot{M}_{\rm He}<-5.93,\\
 1, \hspace{2.0cm}\log\dot{M}_{\rm He}\geq-5.93,\\
\end{array}\right.
\end{equation}
for $1.0 M_{\odot}\leq M_{\rm WD}< 1.1M_{\odot}$.
 \begin{equation}
\eta _{\rm He}=\left\{
 \begin{array}{lc}
 0.54\log\dot{M}_{\rm He}+4.16,\\
   \hspace{2.3cm}\log\dot{M}_{\rm He}<-5.95,\\
-0.54(\log\dot{M}_{\rm He}+5.6)^{\rm 2}+1.01,\\
  \hspace{2.3cm}-5.95\leq\log\dot{M}_{\rm He}<-5.76,\\
 1, \hspace{2.0cm}\log\dot{M}_{\rm He}\geq-5.76,\\
\end{array}\right.
\end{equation}
for $1.1 M_{\odot}\leq M_{\rm WD}< 1.3M_{\odot}$.

 \begin{equation}
\eta _{\rm He}=\left\{
 \begin{array}{lc}
-0.175(\log\dot{M}_{\rm He}+5.35)^{\rm 2}+1.03,\\
  \hspace{2.3cm}\log\dot{M}_{\rm He}<-5.83,\\
 1, \hspace{2.0cm}\log\dot{M}_{\rm He}\geq-5.83,\\
\end{array}\right.
\end{equation}
for $1.3 M_{\odot}\leq M_{\rm WD}< 1.35M_{\odot}$.

 \begin{equation}
\eta _{\rm He}=\left\{
 \begin{array}{lc}
-0.115(\log\dot{M}_{\rm He}+5.7)^{\rm 2}+1.01,\\
  \hspace{2.3cm}\log\dot{M}_{\rm He}<-6.05,\\
 1, \hspace{2.0cm}\log\dot{M}_{\rm He}\geq-6.05,\\
\end{array}\right.
\end{equation}
for $M_{\rm WD}\geq1.35 M_{\odot}$.

We show the total accumulation efficiency, $\eta=\eta_{\rm
He}\eta_{\rm H}$, as a function of mass transfer rate in Fig.
\ref{eta}, which clearly shows the prescription above. We
incorporated this prescription into Eggleton's stellar evolution
code and followed the evolutions of both the mass donor and the
accreting CO WD. The mass lost from the system is assumed to take
away the specific orbital angular momentum of the accreting WD. We
calculated more than 25,000 WD+MS binary systems with various
metallicities, and obtained a large, dense model grid. The initial
masses of donor stars, $M_{\rm 2}^{\rm i}$, range from 0.8
$M_{\odot}$ to 4.3 $M_{\odot}$; the initial masses of CO WDs,
$M_{\rm WD}^{\rm i}$, from 0.629 $M_{\odot}$ to 1.20 $M_{\odot}$;
the initial orbital periods of binary systems, $P^{\rm i}$, from
the minimum value (at which a zero-age main-sequence (ZAMS) star
fills its Roche lobe) to $\sim 20$ days, where the companion star
fills its Roche lobe at the end of the HG. In the calculations, we
assume that the WD explodes as a SN Ia when its mass reaches the
Chandrasekhar mass limit, i.e. 1.378 $M_{\odot}$
(\citealt{NTY84}).

\begin{figure*}
\centerline{\psfig{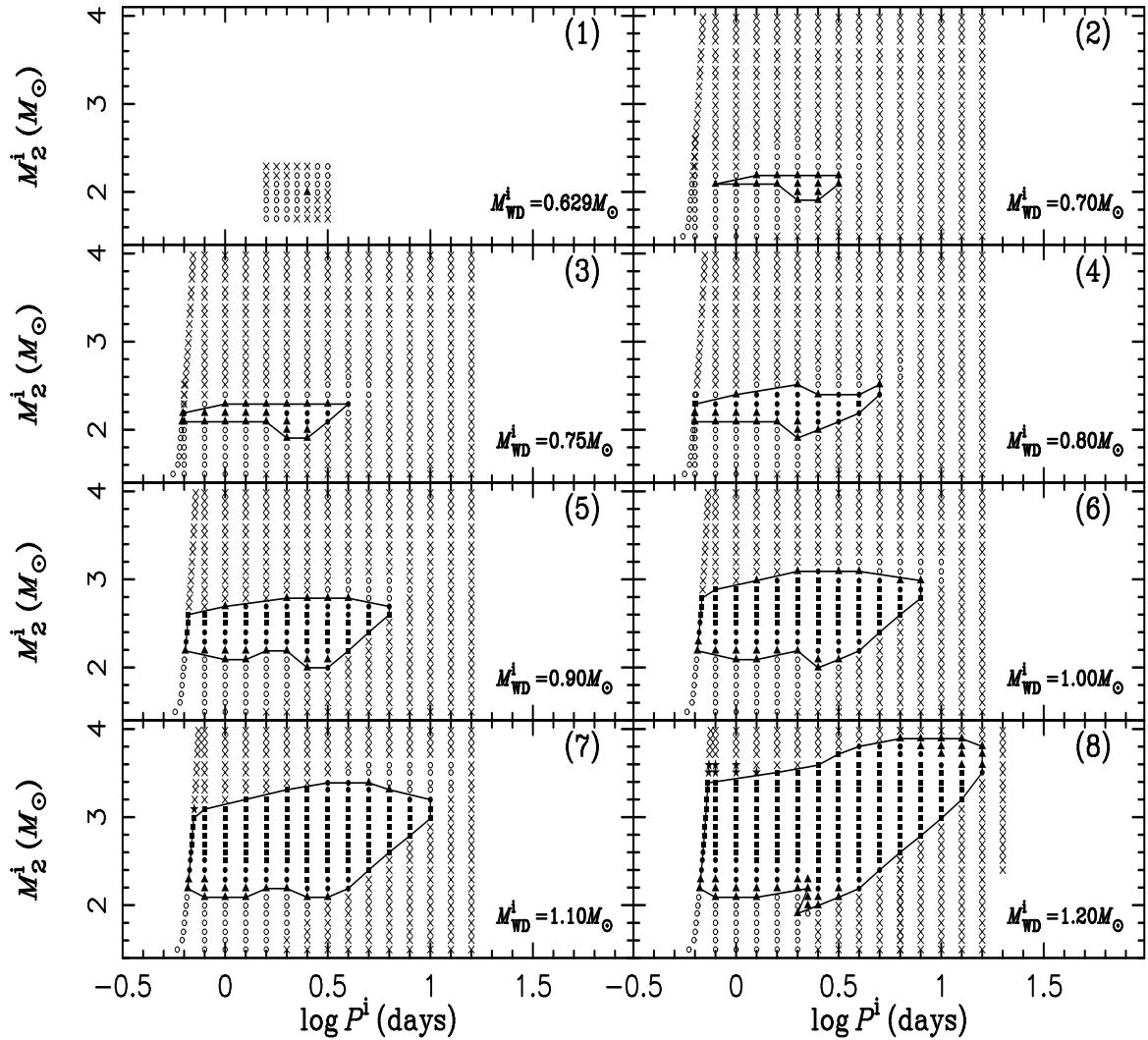}}
\caption{Final outcomes of binary evolution calculation in the
initial orbital period-secondary mass ($\log P^{\rm i}, M_{\rm
2}^{\rm i}$) plane for CO WD +MS binaries with $Z=0.06$, where
$P^{\rm i}$ is the initial orbital period and $M_{\rm 2}^{\rm i}$
is the initial mass of donor star (for different initial WD masses
as indicated in each panel). Filled squares indicate SN Ia
explosions during an optically thick wind phase ($|\dot{M}_{\rm
2}|>\dot{M}_{\rm cr}$). Filled circles denote SN Ia explosions
after the wind phase, where hydrogen-shell burning is stable
($\dot{M}_{\rm cr}\geq |\dot{M}_{\rm 2}|\geq
\frac{1}{2}\dot{M}_{\rm cr}$). Filled triangles denote SN Ia
explosions after the wind phase where hydrogen-shell burning is
mildly unstable ($\frac{1}{2}\dot{M}_{\rm cr}> |\dot{M}_{\rm
2}|\geq \frac{1}{8}\dot{M}_{\rm cr}$). Filled stars denote SN Ia
explosions at delayed dynamical instability phase as shown in
\citet{HAN06}. Open circles indicate systems that experience novae
explosion, preventing the CO WD from reaching 1.378 $M_{\odot}$,
while crosses show the systems that are unstable to dynamical mass
transfer.} \label{fig1}
\end{figure*}

\begin{figure*}
\centerline{\psfig{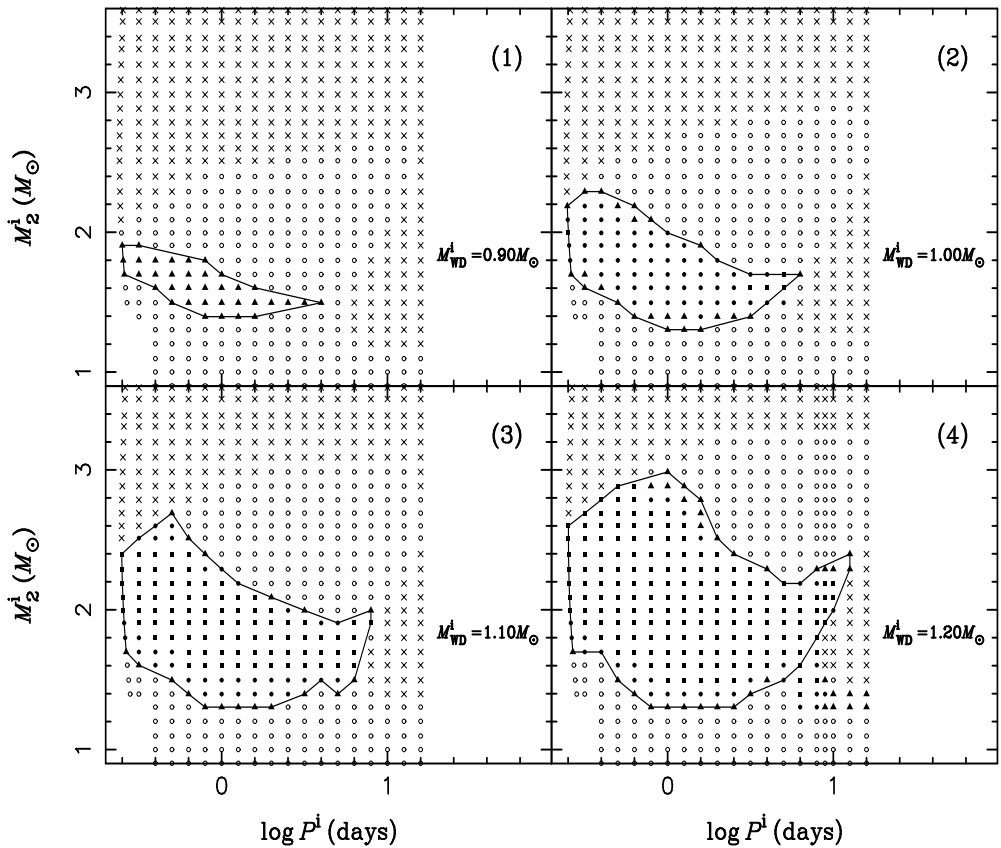}}
\caption{Similar to Fig \ref{fig1} but for $Z=0.0001$. In this
figure, we do not plot the cases with minimum WD mass for
simplicity of appearance of the figure. The minimum WD mass is
$M_{\rm WD}^{\rm min}=0.860 M_{\odot}$ for $Z=0.0001$. Out of the
region surrounded by the solid line in panel (4), there are some
filled symbols representing SNe Ia explosions. We do not include
them in the contour since the mass transfer for these systems
occurs on the giant branch or at the bottom of the giant branch
(see footnote 1). } \label{fig2}
\end{figure*}


\section{BINARY EVOLUTION RESULTS}\label{sect:3}

\begin{figure*}
\centerline{\psfig{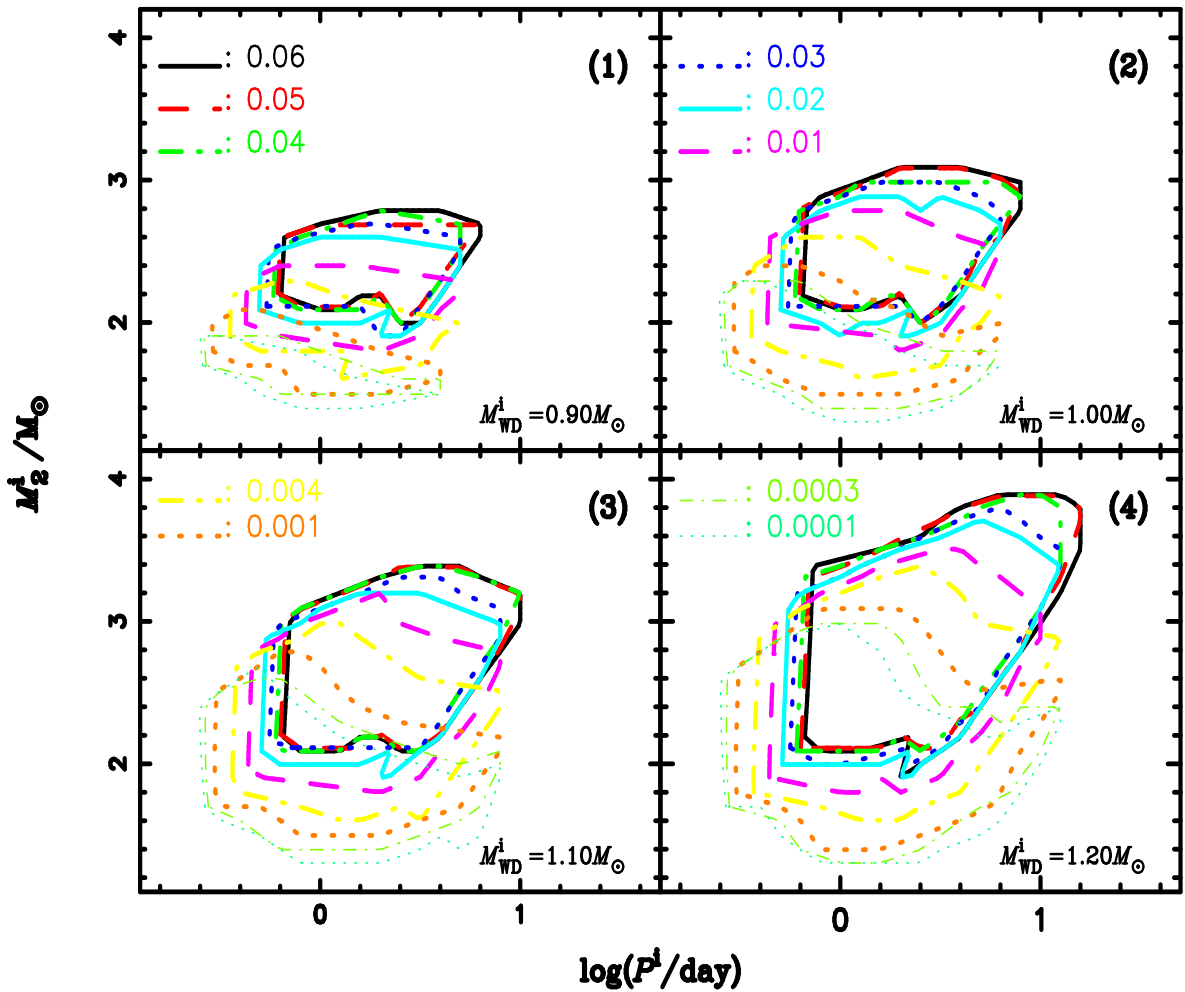}}
\caption{The contours of initial parameters in ($\log P^{\rm
i},M_{\rm 2}^{\rm i}$) for different WD masses and different
metallicities, in which SNe Ia are expected. The initial masses of
WDs are shown in the lower-right region of each panel. Note that,
in all the panels, we use the same types of lines, as indicated,
to represent each metallicity.} \label{120}
\end{figure*}

\subsection{Initial parameters for the progenitor of SNe Ia}\label{subs:3.1}

To conveniently compare our results with previous studies in the
literature, we summarize the final outcomes of all the binary
evolution calculations in the initial orbital period-secondary
mass ($\log P^{\rm i}, M_{\rm 2}^{\rm i}$) plane. For the size
limit of this paper we only show two examples from the ends of the
metallicity range we explored (i.e. $Z=0.06$ and $Z=0.0001$) in
Figs. \ref{fig1} and \ref{fig2}. One can download all the figures
from {\sl http://www.ynao.ac.cn/$^{\rm
\sim}$bps/download/xiangcunmeng.htm}.

From the two figures, we see that CO WDs may reach a mass of 1.378
$M_{\odot}$ during the optically thick wind phase (the filled
squares) or after optically thick wind while in stable (the filled
circles) or unstable (the filled triangles) hydrogen-burning
phase. All these systems are probably progenitors of SNe Ia. Due
to strong hydrogen-shell flash or dynamically unstable mass
transfer, many  CO WDs fail to reach 1.378 $M_{\odot}$. Among
them, some CO WDs have mass larger than 1.30 $M_{\odot}$ at the
onset of dynamically unstable mass transfer (filled stars). These
CO WDs still have a chance to increase their masses to 1.378
$M_{\odot}$ through delayed dynamical instability
(\citealt{HAN06}) and become SNe Ia. These systems are possible
progenitors of the SN 2002ic-like supernovae (\citealt{HAN06}).
The delayed dynamical instability can successfully explain many
properties of SN 2002ic (\citealt{HAN06}), but more evidence is
necessary to confirm this scenario. Thus, we have not included
them in the contours of initial parameter space for SNe Ia (Fig
\ref{fig1}).

For the contours, the left boundaries are determined by the radii
of ZAMS stars, i.e. Roche lobe overflow (RLOF) starts at zero age,
while the systems beyond the right boundaries undergo dynamically
unstable mass transfer at the base of the red giant branch
(RGB)\footnote{Systems consisting of a low-mass secondary and a
high-mass WD possibly undergo dynamically stable mass transfer at
the base of the RGB and even on the RGB because of the low mass
ratio. These systems may also contribute to SNe Ia (WD + RG).
Since the study of WD + RG is beyond the scope of this paper, we
leave out these systems from the contours.}. The upper boundaries
are determined by the delayed dynamical instability (mainly for
$Z>0.004$) and the strong hydrogen-shell flash (mainly for
$Z\leq0.004$). When $Z>0.004$, the systems above the boundaries
have mass ratios too large to stabilize mass transfer. If
$Z\leq0.004$, the mass transfer for the systems above the
boundaries may be dynamically stable, but the mass-transfer rate
is so high that most of the transferred material is lost from the
system by the optically thick wind. The mass-transfer rate then
sharply decreases to less than $\frac{1}{8}\dot{M}_{\rm cr}$ after
mass ratio inversion. The lower boundaries are constrained by the
condition that the mass transfer rate is larger than
$\frac{1}{8}\dot{M}_{\rm cr}$ and that the secondaries have enough
material to transfer onto CO WDs, which can then increase their
masses to 1.378 $M_{\odot}$.

We show the contours in the ($\log P^{\rm i}, M_{\rm 2}^{\rm i}$)
plane for four CO WD masses with various metallicities in Fig.
\ref{120}, from which we may see the influence of metallicity. The
results of other initial CO WD masses are similar to Fig.
\ref{120}. We see that, with the increasing Z, the contour moves
from lower left to upper right in the ($\log P^{\rm i}, M_{\rm
2}^{\rm i}$) plane, indicating that the progenitor systems have
more massive companions and longer orbital periods for a higher Z.
This is due to the correlation between stellar structure and
metallicity. Generally, high metallicity leads to larger radii of
ZAMS stars, then the left hand boundary moves to longer initial
orbital periods. Meanwhile, stars with high metallicity evolve in
a way similar to those with low metallicity but less mass
(\citealt{UME99}; \citealt{CHE07}). Thus, for binaries of CO WDs
with particular orbital periods, the companion mass increases with
metallicity. The situation for the right hand boundaries is a bit
complicated. They move slightly to a longer period with $Z$ when
$Z\geq0.04$, while the tendency is reversed when $Z<0.04$. This
non-monotonic phenomenon is derived from the effect of composition
(determined from equation 1) on opacity and, consequently, on
stellar structure and evolution. For example, with increasing Z,
the radius of a star at the base of RGB decreases when $Z<0.04$,
but increases when $Z\geq0.04$ (\citealt{CHE07}; see
\citealt{MEN07b} for details about the effect of equation (1) on
stellar evolution).

\subsection{The minimum mass of the CO WD for which the WD+MS can produce a SN Ia}\label{subs:3.2}
Previous studies show that the minimum mass of CO WDs leading to SNe
Ia, $M_{\rm WD}^{\rm min}$, may be as low as $0.70 M_{\odot}$ for
$Z=0.02$ (\citealt{LAN00}; \citealt{HAN04}). In our study, $M_{\rm
WD}^{\rm min}$ strongly depends on $Z$, as shown in Fig. \ref{wdm}.
The relation between $M_{\rm WD}^{\rm min}$ and $\log(Z/Z_{\odot})$
is almost linear, and is fitted by

 \begin{equation}
 M_{\rm WD}^{\rm min}/M_{\odot}=0.6679-0.08799\log(Z/Z_{\odot}),
  \end{equation}
where the relative error of $M_{\rm WD}^{\rm min}$ is less than
2.3\% in the fitting. We see in Fig. \ref{wdm} that the minimum
mass sharply decreases with metallicity and the difference in
$M_{\rm WD}^{\rm min}$ between the cases of $Z=0.06$ and
$Z=0.0001$ may be up to 0.24 $M_{\odot}$. We explain this as
follows: As mentioned in subsection \ref{subs:3.1}, for a high
$Z$, the companions in WD + MS systems for SNe Ia are more
massive, and so have more material to transfer onto CO WDs. The CO
WDs therefore do not need to be as massive for the production of
SNe Ia. Meanwhile, the time-scale for mass transfer is the thermal
time-scale, which increases with metallicity. This leads to a
lower mass-transfer rate for a high metallicity. As a consequence,
less material is lost via optically thick wind. This means that a
larger fraction of material can be accumulated on the surfaces of
CO WDs for systems with a high Z. Therefore CO WDs can be
initially less massive for SNe Ia for high Z.

\section{Binary population synthesis}\label{sect:4}
Adopting the results in section \ref{sect:3},  we have studied the
supernova frequency from the WD+MS channel via the rapid binary
evolution code developed by \citet{HUR00, HUR02}. Here after, we
use {\sl primordial} to represent the binaries before the
formation of WD+MS systems and {\sl initial} for WD+MS systems.

\subsection{Common envelope in binary evolution}\label{subs:4.1}
During binary evolution, the primordial mass ratio (primary to
secondary) is crucial for the first mass transfer. If it is larger
than a critical mass ratio, $q_{\rm c}$, the first mass transfer
is dynamically unstable and a common envelope (CE) forms
(\citealt{PAC76}). \textbf{The ratio} $q_{\rm c}$ varies with the
evolutionary state of the primordial primary at the onset of RLOF
(\citealt{HW87}; \citealt{WEBBINK88}; \citealt{HAN02};
\citealt{POD02}; \citealt{CHE08}). In this study, we adopt $q_{\rm
c}$ = 4.0 when the primary is on MS or HG. This value is supported
by detailed binary evolution studies (\citealt{HAN00};
\citealt{CHE02, CHE03}). If the primordial primary is on FGB or
AGB, we use

\begin{equation}
q_{\rm c}=[1.67-x+2(\frac{M_{\rm c1}^{\rm P}}{M_{\rm 1}^{\rm
P}})^{\rm 5}]/2.13,  \label{eq:qc}
  \end{equation}
where $M_{\rm c1}^{\rm P}$ is the core mass of primordial primary,
and $x={\rm d}\ln R_{\rm 1}^{\rm P}/{\rm d}\ln M_{\rm 1}^{\rm p}$
is the mass-radius exponent of primordial primary and varies with
composition. If the mass donors (primaries) are naked helium
giants, $q_{\rm c}$ = 0.748 based on equation (\ref{eq:qc}) (see
\citealt{HUR02} for details).

Embedded in the CE is a ``new'' binary consisting of the dense
core of the primordial primary and the primordial secondary. Due
to frictional drag within the envelope, the orbit of the ``new''
binary decays and a large part of the orbital energy released in
the spiral-in process is injected into the envelope
(\citealt{LS88}). Here, we assume that the CE is ejected if
\begin{equation} \alpha_{\rm CE}\Delta E_{\rm orb}=|E_{\rm
bind}|,
  \end{equation}
where $\Delta E_{\rm orb}$ is the orbital energy released, $E_{\rm
bind}$ is the binding energy of \textbf{CE}, and $\alpha_{\rm CE}$
is CE ejection efficiency (i.e. the fraction of the released
orbital energy used to eject the CE). This criterion is known as
standard $\alpha$-formalism (\citealt{NELEMANS00};
\citealt{NELEMANS05}). Since the internal energy in the envelope
is not incorporated into the binding energy, $\alpha_{\rm CE}$ may
be greater than 1 (see \citealt{HAN95} for details about the
internal energy). In this paper, we set $\alpha_{\rm CE}$ to be
1.0 or 3.0.

\citet{NELEMANS00} and \citet{NELEMANS05} found that it is
difficult for the standard $\alpha$-formalism to reproduce a close
pair of white dwarfs without the inclusion of the thermal energy
of the envelope. They then suggested an alternative algorithm
equating the angular momentum balance (the $\gamma$-algorithm,
\citealt{NELEMANS05}), which may explain the formation of all
kinds of close binaries. However, \citet{WEBBINK07} argued that
the $\alpha$-formalism with envelope internal energy included can
reproduce observations best. Note that the formation of WD + MS
binaries is not much affected by the use of the $\gamma$-algorithm
in comparison to $\alpha$-formalism, especially when $\alpha_{\rm
CE}=3.0$ (\citealt{NELEMANS05}). Therefore, we continue to use the
$\alpha$-formalism throughout this paper and the main results in
this paper will hold regardless.

\begin{figure}
\centerline{\psfig{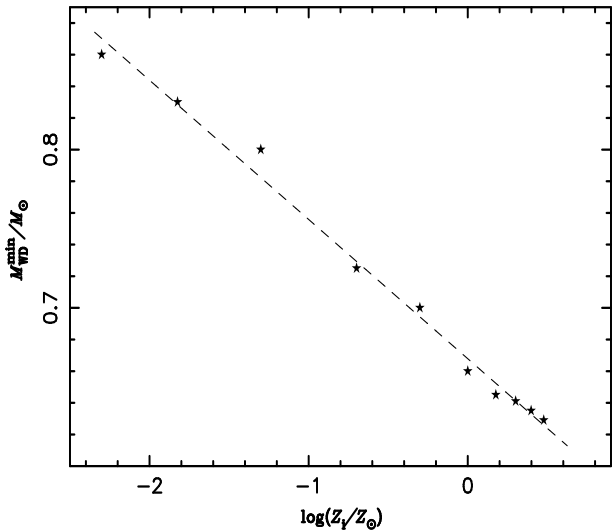}}
\caption{Relation between the minimum mass of CO WD for the
production of SNe Ia and metallicity, with which the CO WD can
increase its mass to 1.378 $M_{\rm \odot}$.} \label{wdm}
\end{figure}

\subsection{Evolutionary channels to WD + MS systems}\label{subs:4.2}
According to the evolutionary phase of the primordial primary at the
onset of the first RLOF, WD + MS systems can be produced from three
channels.

Case 1 (He star channel): the primordial primary is in HG or
\textbf{on} RGB at the onset of the first RLOF (i.e. case B
evolution defined by \citealt{KW67}). In this case, a CE is formed
because of a large mass ratio or a convective envelope of the mass
donor. After the CE ejection (if it occurs), the mass donor
becomes a helium star and continues to evolve. The helium star
likely fills its Roche lobe again after the central helium is
exhausted. Since the mass donor is much less massive than before,
this RLOF is dynamically stable, resulting in a close CO WD+MS
system (see \citealt{NOM99, NOM03} for details). The primordial
orbital periods may range from 10 to 1000 days for this channel.

Case 2 (EAGB channel): the primordial primary is in early
asymptotic giant branch stage (EAGB) (i.e. helium is exhausted in
the core, while thermal pulses have not yet started). A CE is
formed because of dynamically unstable mass transfer. After the CE
is ejected, the orbit decays and the primordial primary becomes a
helium red giant (HeRG). The HeRG may fill its Roche lobe and
start the second RLOF. Similar to the He star channel, this RLOF
is stable and produces WD + MS systems after RLOF. The range of
the primordial periods is $\sim$ 100 -- 1000 days for this
channel. During the EAGB, the hydrogen-burning shell is extinct in
the primary, which has the same core mass as that at the bottom of
the EAGB (\citealt{HUR00}).

Case 3 (TPAGB channel): the primordial primary fills its Roche lobe
at the thermal pulsing AGB (TPAGB) stage. Similar to the above two
channels, a CE is formed during the RLOF. A CO WD + MS binary is
produced after CE ejection. The primordial periods of the systems
experiencing the channel are larger than 1000 days.

Generally, the CO WD + MS binaries obtained from the different
channels above have some different properties and these properties
eventually determine whether the systems may explode as SNe Ia or
not. We will discuss this in section \ref{sect:5}. The WD + MS
systems continue to evolve and the MS companions may fill their
Roche lobes, and transfer their material to the CO WDs, which are
likely to explode as SNe Ia. Here, we assume that, if the initial
orbital period, $P_{\rm orb}^{\rm i}$, and the initial secondary
mass, $M_{\rm 2}^{\rm i}$, of a WD + MS system is located in the
appropriate regions in the ($\log P^{\rm i}, M_{\rm 2}^{\rm i}$)
plane (see Figs. \ref{fig1} to \ref{fig2}) for SNe Ia at the onset
of RLOF, a SN Ia is produced.

\subsection{Basic parameters for Monte Carlo simulations}\label{subs:4.3}
To investigate the birth rate of SNe Ia, we followed the evolution
of $10^{\rm 7}$ binaries for various $Z$ via Hurley's rapid binary
evolution code (\citealt{HUR00, HUR02}). The results of grid
calculations in section \ref{sect:3} are incorporated into the
code. The code is only valid for $Z\leq0.03$, and the production
of SNe Ia of seven metallicities, i.e. $Z=0.03$, $0.02$, $0.01$,
$0.004$, $0.001$, $0.0003$ and $0.0001$, are investigated here.
The primordial binary samples are generated in a Monte Carlo way
and a circular orbit is assumed for all binaries. The basic
parameters for the simulations are as follows.

(i) The initial mass function (IFM) of \citet{MS79} is adopted.
The primordial primary is generated according to the formula of
\citet{EGG89}
\begin{equation}
M_{\rm 1}^{\rm p}=\frac{0.19X}{(1-X)^{\rm 0.75}+0.032(1-X)^{\rm
0.25}},
  \end{equation}
where $X$ is a random number uniformly distributed in the range
[0,1] and $M_{\rm 1}^{\rm p}$ is the mass of the primordial primary,
which ranges from 0.1 $M_{\rm \odot}$ to 100 $M_{\rm \odot}$.

(ii) The mass ratio of the primordial binaries, $q'$, is a very
important parameter for binary evolution. For simplicity, we take a
uniform mass-ratio distribution (\citealt{MAZ92}; \citealt{GM94}):
\begin{equation}
n(q')=1, \hspace{2.cm} 0<q'\leq1,
\end{equation}
where $q'=M_{\rm 2}^{\rm p}/M_{\rm 1}^{\rm p}$.

(iii) We assume that all stars are members of binary systems and
that the distribution of separations is constant in $\log a$ for
wide binaries, where $a$ is separation, and falls off smoothly at
small separation:
\begin{equation}
a\cdot n(a)=\left\{
 \begin{array}{lc}
 \alpha_{\rm sep}(a/a_{\rm 0})^{\rm m} & a\leq a_{\rm 0};\\
\alpha_{\rm sep}, & a_{\rm 0}<a<a_{\rm 1},\\
\end{array}\right.
\end{equation}
where $\alpha_{\rm sep}\approx0.070$, $a_{\rm 0}=10R_{\odot}$,
$a_{\rm 1}=5.75\times 10^{\rm 6}R_{\odot}=0.13{\rm pc}$ and
$m\approx1.2$. This distribution implies that the numbers of wide
binary systems per logarithmic interval are equal, and that
approximately 50 percent of the stellar systems have orbital periods
less than 100 yr (\citealt{HAN95}).

(iv)We simply assume a single starburst (i.e. $10^{\rm 11}
M_{\odot}$ in stars are produced one time) or a constant star
formation rate $S$ (SFR) over the last 15 Gyr calibrated so that
one binary with $M_{\rm 1}>0.8 M_{\odot}$ is born in the Galaxy
each year (see \citealt{IT84}; \citealt{HAN95}; \citealt{HUR02}).
From this calibration, we can get $S=5$ ${\rm yr^{-1}}$ (see also
\citealt{WK04}). The constant star formation rate is consistent
with the estimation of \citet{TIM97}, which successfully
reproduces the $^{26}$Al 1.809-MeV gamma-ray line and the
core-collapse supernova rate in the Galaxy (\citealt{TIM97}).

\section{The results of binary population synthesis}\label{sect:5}

\begin{figure}
\centerline{\psfig{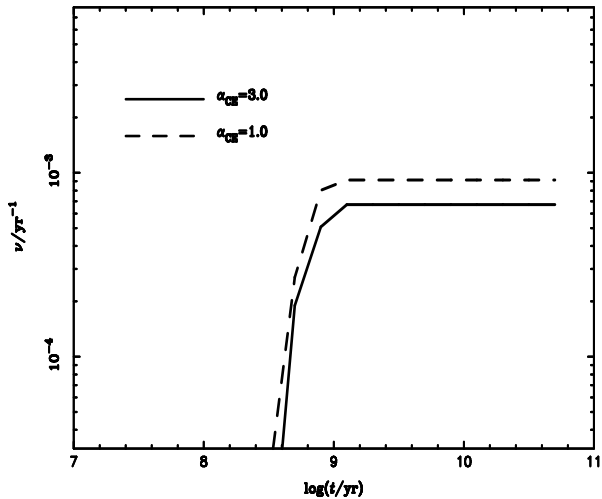}}
\caption{The evolution of the birth rates of SNe Ia for a constant
star formation rate (Z=0.02, SFR=$5 M_{\rm \odot}{\rm yr^{\rm
-1}}$). Solid and dashed lines show the cases with $\alpha_{\rm
CE}=3.0$ and $\alpha_{\rm CE}=1.0$, respectively.} \label{sfrbirth}
\end{figure}

\begin{figure}
\centerline{\psfig{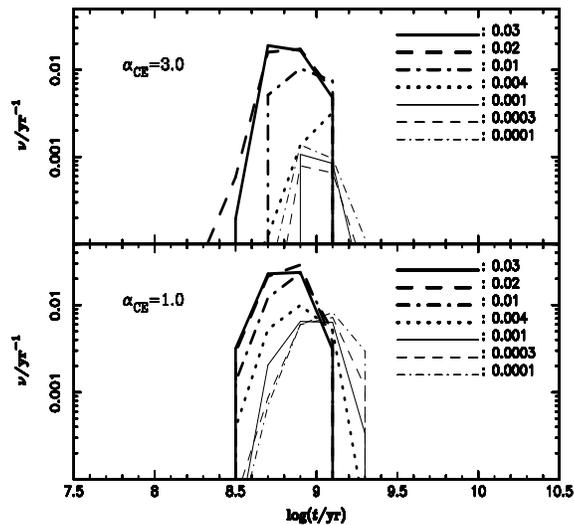}}
\caption{The evolution of the birth rates of SNe Ia for a single
starburst of $10^{\rm 11}M_{\odot}$ for different metallicities. The
upper panel shows the cases with $\alpha_{\rm CE}=3.0$,  while the
cases of $\alpha_{\rm CE}=1.0$ are in the bottom panel.}
\label{birth}
\end{figure}
\subsection{the Birth rates of SNe Ia}\label{subs:5.1}
Fig. \ref{sfrbirth} shows the Galactic birth rates of SNe Ia (i.e.
$Z=0.02$ and SFR= 5.0$M_{\odot}/{\rm yr}$) from the WD+MS channel.
From the figure, we see that the Galactic birth rate is around
0.7-1.0$\times10^{\rm -3}{\rm yr^{\rm -1}}$, consistent with that
of \citet{HAN04}. This result is lower but comparable to that
inferred from observations (3-4$\times10^{\rm -3}{\rm yr^{\rm
-1}}$, \citealt{VAN91}; \citealt{CT97}).

The birth rate of SNe Ia at various metallicities are presented in
Fig. \ref{birth}. In this figure, we see that most supernovae
occur between 0.2 Gyr and 2 Gyr after the starburst. An
interesting phenomenon is that a high metallicity leads to a
systematically earlier explosion time of SNe Ia, which due to the
effect of metallicity $Z$ on the maximum initial mass of the
companion for SNe Ia and on the stellar evolution. As shown in
Fig. \ref{120}, $M_{\rm 2}^{\rm i}$ increases with metallicity
$Z$. Generally, a massive star evolves more quickly than a
low-mass one. So, the explosion time is earlier with $Z$ from this
view. Though a high $Z$ also slows the evolution of a star down,
its influence is much less than that of stellar mass based on
detailed calculations of stellar evolution (\citealt{UME99};
\citealt{CHE07}).

We also see in Fig. \ref{birth} that the peak value increases with
metallicity $Z$. This comes from the fact that the parameter space
for SNe Ia increases with metallicity, e.g. the range of the
initial masses of WDs is larger for a high metallicity. In the
upper panel of Fig. \ref{birth}, however, the peak value for the
case of $Z=0.0001$ is higher than that of $Z=0.001$, which is
mainly from the influence of $\alpha_{\rm CE}$ and $Z$ on the
TPAGB channel (see subsection \ref{subs:5.2.1}). $\alpha_{\rm CE}$
significantly affects the peak values only at low metallicities,
that is, the peak value of the birth rate is obviously larger for
$\alpha_{\rm CE}=1.0$ in comparison to $\alpha_{\rm CE}=3.0$ when
$Z<0.004$. These phenomena originate from the influences of
$\alpha_{\rm CE}$ and metallicity on the He star channel and the
TPAGB channel (see subsection \ref{subs:5.2.1} for details).

However, the results above are opposite to the recent study of
\citet{GUO08}, where an analytic method was used (see also
\citealt{HAC99a, HAC99b} and \citealt{NOM99, NOM03}).
\citet{GUO08} found that the peak of the birth rate for a low $Z$
appears significantly earlier than that for a high $Z$. Also, they
have not found the dependence of the peak value of birth rate on
metallicity. This discrepancy comes from the different approaches
in calculating $|\dot{M}_{\rm 2}|$ in the two studies. As
described in section \ref{sect:2}, the value of $|\dot{M}_{\rm
2}|$ is crucial for the accretion of CO WDs. Thus, initial
parameter spaces for SNe Ia are different in the two studies, i.e.
a low metallicity leads to a higher maximum mass of the
progenitors in \citet{GUO08}. This tendency is the opposite of
that in this paper. Then, the delay times of SNe Ia, which are
closely relevant to the initial companion mass, are also
different. The final birth rates are therefore different. For
example, given $Z=0.02$, the peak value from the analytic method
is larger than that from detailed binary evolution calculations by
a factor of 3, if a single starburst is assumed (\citealt{HAN04}).
Note that our results are more physical than those obtained from
the simple analytic method used by \citet{HAC99a, HAC99b},
\citet{NOM99, NOM03} and \citet{GUO08}, as our results are based
on detailed binary evolution calculations with the latest input
physics.

\begin{figure*}
\centerline{\psfig{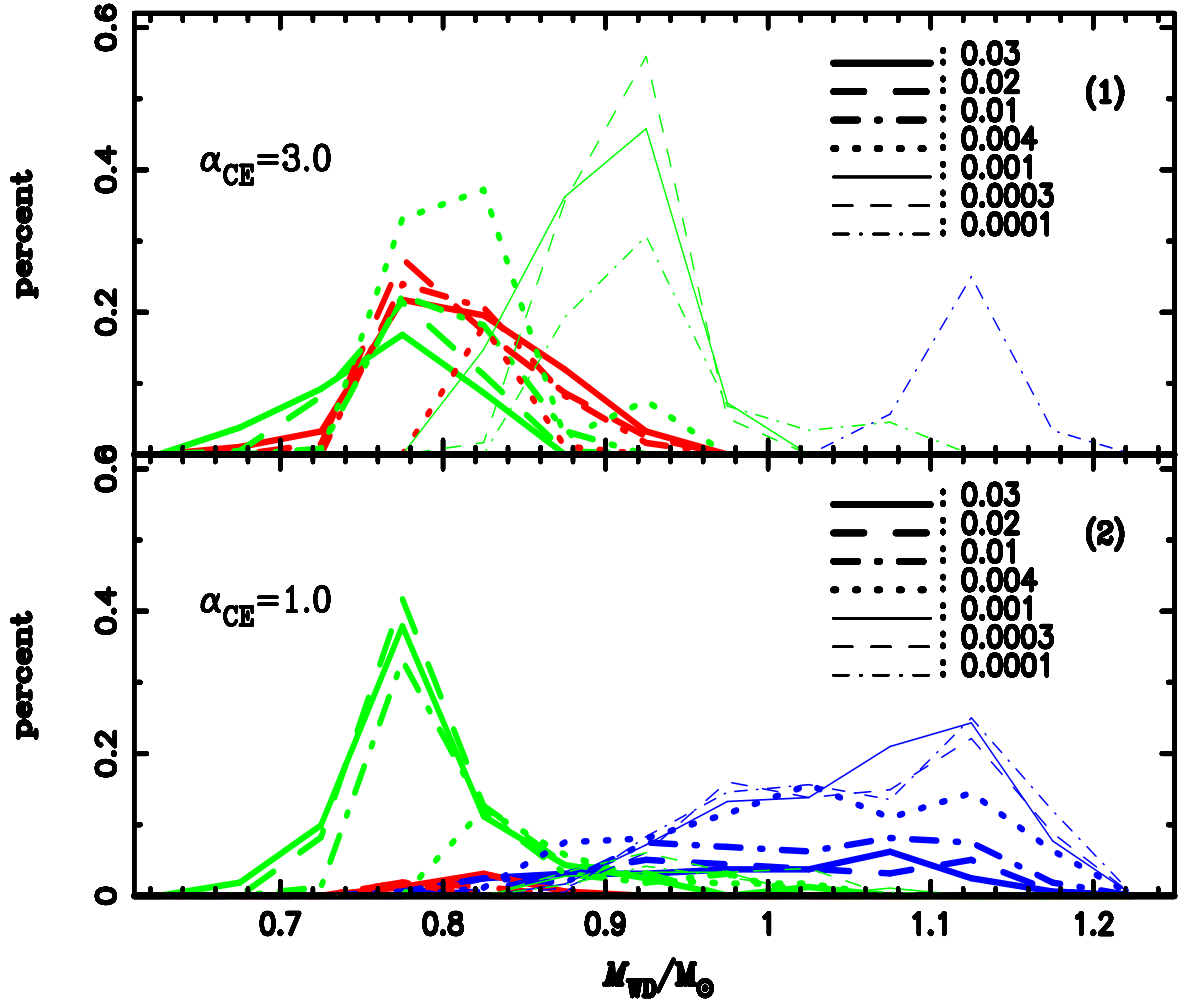}}
\caption{The distribution of the initial WD masses in WD + MS
systems from different SN Ia channels for various metallicities
and different $\alpha_{\rm CE}$. The red, green and blue lines
represent those from He star, EAGB and TPAGB channel,
respectively.} \label{wddistong}
\end{figure*}

\subsection{Distribution of initial parameters of WD + MS systems for SNe Ia}\label{subs:5.2}
Observationally, some WD + MS systems are possible progenitors of
SNe Ia (see the review of \citealt{PAR07}). Further studies are
necessary to finally confirm them (from both observations and
theories). In this section, we will present some properties of
initial WD + MS systems for SNe Ia , which may help us to search
for the potential progenitors of SNe Ia in various environments.

\subsubsection{Distribution of initial masses of WDs}\label{subs:5.2.1}
The distribution of initial WD masses from different channels are
shown in Fig. \ref{wddistong}, in which the relative importance of
the three channels is clearly shown. In Fig. \ref{mwddis}, we
present the distribution of the initial masses of the CO WDs
leading to SNe Ia for various $Z$ and $\alpha_{\rm CE}$. Fig.
\ref{wddistong} shows that $\alpha_{\rm CE}$ is crucial for the He
star channel. If a system experiences CE phase before rather than
after helium burning, it will more likely merge due to a large
binding energy and a short primordial orbital period, especially
for $\alpha_{\rm CE}=1.0$. Thus, the contribution of WD + MS
systems from the He star channel \textbf{is} less when
$\alpha_{\rm CE}=1.0$ and increases when $\alpha_{\rm CE}=3.0$.
Since the stellar radius increases with metallicity $Z$, a
primordial binary with given $M_{\rm 1}^{\rm P}$, $M_{\rm 2}^{\rm
P}$ and $P^{\rm P}$ is more likely to undergo He star channel
evolution if it has a high $Z$. The importance of the He star
channel thus increases with metallicity $Z$. Actually, no WD + MS
systems result from this channel when $Z<0.004$ in our simulation.
He star channel has an important contribution to WD + MS systems
with CO WDs around $0.78 M_{\odot}$ (see the peaks at low masses
in Fig. \ref{mwddis}).

The EAGB channel \textbf{is not} significantly influenced by
$\alpha_{\rm CE}$ for its moderate primordial orbital period, and
it is the only contributor to the peaks at masses$\sim 0.92
M_{\odot}$ in the upper panel of Fig. \ref{mwddis}. EAGB channel
also produces many WD + MS system with CO WD mass around $0.78
M_{\odot}$ (see Fig \ref{wddistong}).

The massive CO WDs (i.e the high-mass tail for $Z>0.004$, the
plateau from $0.9 M_{\odot}$ to $1.20 M_{\odot}$ for $Z\leq0.004$
with $\alpha_{\rm CE}=1.0$, and peak around $1.13 M_{\odot}$ with
$\alpha_{\rm CE}=3.0$ and $Z=0.0001$ in Fig. \ref{mwddis}) are
mainly from TPAGB channel (see the blue lines in Fig.
\ref{wddistong}). Because of the low binding energy of the common
envelope and a long primordial orbital period, $\alpha_{\rm CE}$
has a remarkable influence on CO+WD systems from the TPAGB
channel. Generally, if a CE can be ejected, a low $\alpha_{\rm
CE}$ produces a shorter orbital-period WD + MS system, which is
more likely to fulfill the conditions for SNe Ia. Therefore, we
see obvious contributions from the TPAGB channel when $\alpha_{\rm
CE}=1.0$, but no contribution from this channel when $\alpha_{\rm
CE}=3.0$ except for the case of $Z=0.0001$. We explain this as
follows: For a low $Z$, due to relatively small stellar radius,
binaries undergoing TPAGB channels usually have shorter primordial
orbital periods and tighter common envelopes, and the produced CO
WD + MS binaries have shorter orbital periods after CE ejection
and are more likely to contribute to SNe Ia. This effect increases
with the decrease of metallicity. We therefore see the high-mass
peak in the figure for metallicity as low as 0.0001 with
$\alpha_{\rm CE}=3.0$. It is a little bit different with
$\alpha_{\rm CE}=1.0$, for which the produced WD + MS systems have
shorter initial orbital periods in comparison to $\alpha_{\rm
CE}=3.0$. Thus, many WD + MS systems, which can not lead to SNe Ia
with $\alpha_{\rm CE}=3.0$, contribute to SNe Ia with $\alpha_{\rm
CE}=1.0$, and produce a plateau of CO WD masses instead of a peak
(the bottom panel in Fig. \ref{mwddis}).

As discussed above, for high $Z$, the contribution to SNe Ia from
the He star channel increases, while that from the TPAGB channel
decreases with $\alpha_{\rm CE}$, which leads to a nearly constant
birth rate of SNe Ia for different values of $\alpha_{\rm CE}$.
However, for the cases with low metallicities, the He star channel
has no contribution to SNe Ia at either value adopted for
$\alpha_{\rm CE}$, while a large amount of WD + MS systems from
the TPAGB channel contribute to SNe Ia at the low $\alpha_{\rm
CE}$, resulting in a higher birth rate in comparison to the high
$\alpha_{\rm CE}$.



\subsubsection{Distribution of initial secondary masses}\label{subs:5.2.2}
Fig. \ref{mmsdis} presents the distributions of the initial masses
of secondaries for SNe Ia for various metallicity $Z$ and
$\alpha_{\rm CE}$. The distributions for different metallicities
have similar shapes (i.e. a low-mass sharp peak with a high-mass
tail). Since the contours for SNe Ia move to higher masses with
$Z$, the peak mass of the secondaries also moves to higher masses
with $Z$. The difference of the peak initial masses of the
secondaries between $Z=0.03$ and $Z=0.0001$ is as large as
$\sim0.7 M_{\rm \odot}$.

\subsubsection{Distribution of initial orbital periods}\label{subs:5.2.3}
The distributions of the initial orbital periods are shown in Fig.
\ref{perdis}. Similar to the distribution of initial masses of CO
WDs, there are also double peaks for $Z=0.0001$ and $\alpha_{\rm
CE}=3.0$, which correspond to the low-mass peak and the high-mass
one in Fig. \ref{mwddis}, respectively.

\subsubsection{Distribution of initial separations}\label{subs:5.2.4}
The distributions of the initial separations are shown in Fig.
\ref{sepdis}. Similar to the distributions of the initial masses
of CO WDs and that of the initial orbital periods, there are
double peaks for the case of $Z=0.0001$ and $\alpha_{\rm CE}=3.0$.
The low-separation peak is from EAGB channel and the
high-separation one is from TPAGB channel.

We see from \textbf{Fig. \ref{sepdis}} that there are almost no
initial systems with separations larger than $\sim 26 R_{\rm
\odot}$. This is a natural result since we only focus on WD + MS
systems in this paper. For systems with long initial separations,
mass transfer occurs on RGB, which is beyond the scope of this
paper.

\begin{figure}
\centerline{\psfig{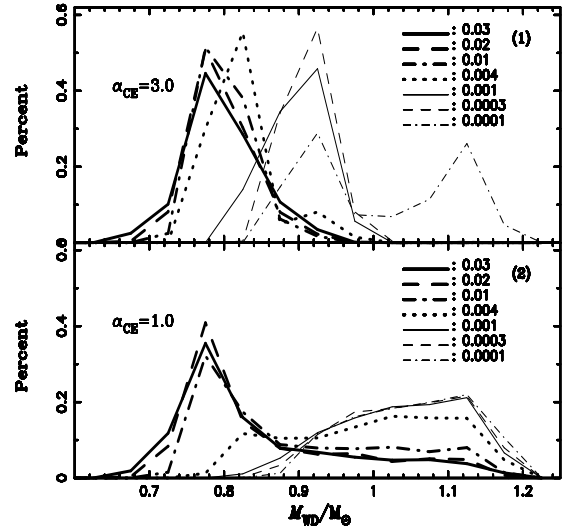}}
\caption{The distribution of the initial WD masses in WD + MS
systems which can ultimately produce SNe Ia for various
metallicities and different $\alpha_{\rm CE}$.} \label{mwddis}
\end{figure}

\begin{figure}
\centerline{\psfig{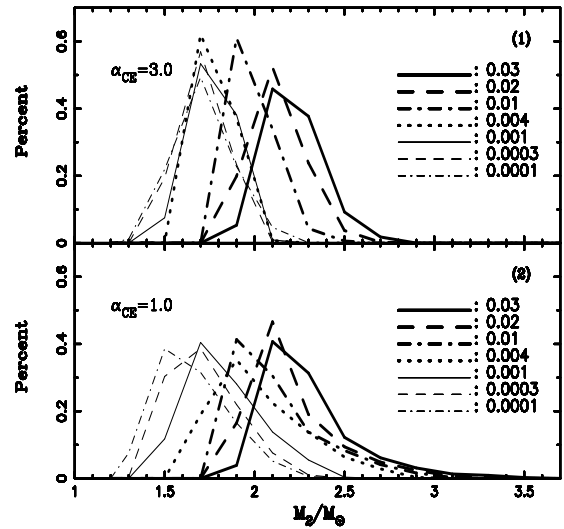}}
\caption{The distribution of the initial masses of secondaries in WD
+ MS systems which can ultimately produce SNe Ia for different
metallicities and different $\alpha_{\rm CE}$.} \label{mmsdis}
\end{figure}

\begin{figure}
\centerline{\psfig{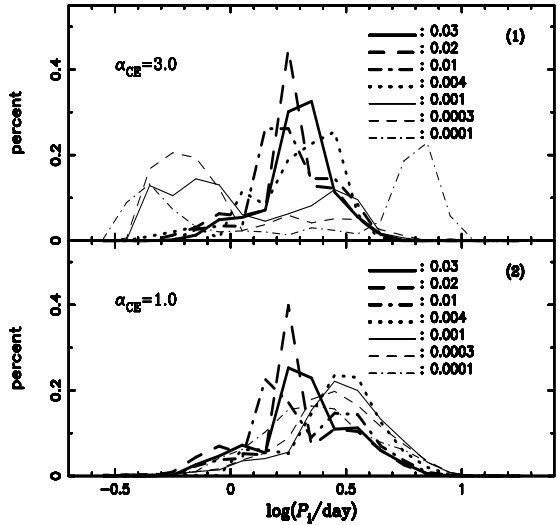}}
\caption{The distribution of the initial orbital periods of the
WD+MS systems which can ultimately produce SNe Ia for different
metallicities and different $\alpha_{\rm CE}$.} \label{perdis}
\end{figure}

\begin{figure}
\centerline{\psfig{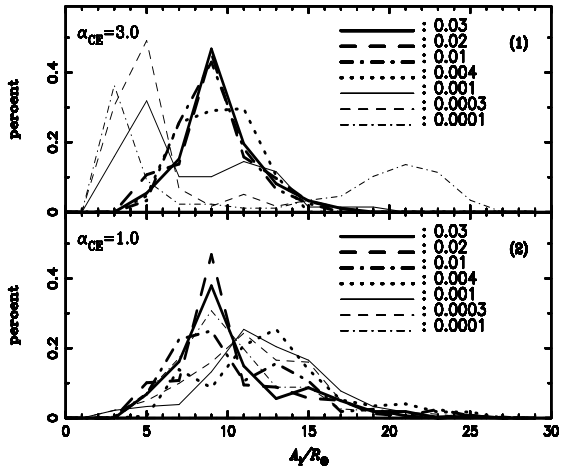}}
\caption{The distribution of the initial separations of the WD+MS
systems which can ultimately produce SNe Ia for different
metallicities and different $\alpha_{\rm CE}$.} \label{sepdis}
\end{figure}
\section{discussion}\label{sect:6}
\subsection{Correlation between the peak luminosity of SNe Ia and metallicity}\label{subs:6.1}
In our study, the masses of CO WDs leading to SNe Ia are more
massive in a low-metallicity environment (see Figs \ref{wdm} and
\ref{mwddis}). Some previous studies showed that a massive CO WD
leads to a lower C/O ratio, and thus a lower amount of $^{\rm
56}{\rm Ni}$ synthesized in the thermonuclear explosion
(\citealt{NOM99, NOM03}), which results in a lower luminosity of
SNe Ia (\citealt{ARN82}; \citealt{ARN85}; \citealt{BRA92}).
According to this, our results means that the mean maximum
luminosity of SNe Ia increases with metallicity $Z$, implying that
the average maximum luminosity decreases with redshift, since
metallicity generally decreases with redshift. Observationally,
the mean peak brightness of SNe Ia in a galaxy has less variation
in the outer regions than in the inner regions (\citealt{WAN97};
\citealt{RIE99}). \citet{NOM99, NOM03} suggested that this results
from the effect of metallicity, i.e. the maximum peak brightness
of SNe Ia is larger in the inner region than that in the outer
region of a galaxy since metallicity decreases along the radial
direction of a galaxy (\citealt{BAES07}; \citealt{LEMASLE07}),
while the minimum peak brightness of SNe Ia is similar because the
maximum initial CO WD mass is almost equal, i.e. $\sim 1.2
M_{\odot}$ (see Fig. \ref{mwddis} in this paper and Fig. 1 in
\citealt{MEN07b}), which leads to the lowest C/O ratio of a CO WD
and thus the least amount of $^{\rm 56}{\rm Ni}$ synthesized in
the thermonuclear explosion (\citealt{NOM99, NOM03}). In addition,
\citet{HAM96b} did not find the classical Malmquist bias in their
sample and the most distant SNe Ia are not significantly brighter
than nearby samples. This phenomenon is at least not inconsistent
with our prediction.

\citet{LAN00} suggested that on average, the initial masses of CO
WDs for SNe Ia with a low metallicity ($Z=0.001$) is larger than
that with a high metallicity ($Z=0.02$) by about 0.2 $M_{\rm
\odot}$. From the BPS study in this paper, the difference of the
CO WD masses from various metallicities may be up to 0.3
$M_{\odot}$ (see Fig \ref{mwddis}), consistent with the suggestion
of \citet{LAN00}. Detailed single stellar evolution calculations
(\citealt{HAN94}; \citealt{GIRARDI00}; \citealt{MEN07b}) show
that, for a given main-sequence star, the CO WD mass decreases
with metallicity Z, that is, a 0.2 $M_{\rm \odot}$ difference of
CO WD masses can be obtained from different metallicities
(\citealt{MEN07b}). Since the CO WD masses from single stellar
evolution may be taken as upper limits of CO WD masses from binary
evolution to some extent, the mass difference of CO WDs for SNe Ia
from different metallicties might be achieved from the evolution
prior to CO WD formation.

\subsection{Other possible channels for SNe Ia}\label{subs:6.2}
From our study, the WD + MS channel can only account for about
$1/3$ of the SNe Ia observed(\citealt{VAN91}; \citealt{CT97}).
Therefore, there may be other channels or mechanisms contributing
to SNe Ia. As mentioned in section \ref{sect:1}, a wide symbiotic
system, WD + RG, is a possible progenitor of SNe Ia
(\citealt{HAC99b}), though previous BPS studies indicated little
contributions from this channel (\citealt{YUN98};
\citealt{HAN04}). Recently, \citet{HK05, HK06a, HK06b} and
\citet{HKL07} suggested that several recurrent novae are probable
progenitors of SNe Ia and some of them belong to the WD+RG
channel. Meanwhile, \citet{PAT07} suggested that the companion of
the progenitor of SN 2006X may be an early RGB star. So, a further
study of this channel is necessary. An alternative is the
double-degenerate (DD) channel (\citealt{IT84}; \citealt{WI87}),
although it is theoretically less favored (\citealt{HN00}). In
this channel, two CO WDs with a total mass larger than the
Chandrasekhar mass limit may coalesce and explode as a SN Ia. The
birth rate from this channel is comparable to the observational
value (\citealt{HAN98}; \citealt{YUN98, YUN00}; \citealt{TUT02}),
and SN 2003fg and SN 2005hj likely resulted from the DD channel
(\citealt{HOW06}; \citealt{BRA06}; \citealt{QUI07}).
Observationally, a large amount of DD systems are discovered
(\citealt{NAPIWOTZKI04}), but only KPD 1930+2752 is a possible
progenitor candidate for a SN Ia via DD channel (\citealt{GEI07}).
The total mass of KPD 1930+2752 ($\sim1.52 M_{\rm \odot}$) exceeds
the Chandrasekhar mass limit and the time scale of coalescence is
about 200 Myr estimated from orbital shrinkage caused by
gravitational wave radiation (\citealt{GEI07}). However,
\citet{ERGMA01} argued that, from detailed binary evolution
calculation, the final mass of KPD 1930+2752 is smaller than the
Chandrasekhar mass limit due to a large amount of mass loss during
evolution. In addition, earlier numerical simulations showed that
the most probable fate of the coalescence is an accretion-induced
collapse and, finally,  neutron star formation (see the review by
\citealt{HN00}). A definitive conclusion for DD model is thus
premature at present, and further studies are needed.

\citet{LIE03, LIE05} and \citet{WF05} found that about 10\% of WDs
have magnetic fields higher than 1MG. The mean mass of these WDs
is 0.93 $M_{\odot}$, compared to mean mass of all WDs which is
0.56 $M_{\odot}$ (see the review by \citealt{PAR07} for details).
{Thus, the magnetic WDs are more likely to reach the Chandrasekhar
mass limit by accretion. Meanwhile, the magnetic field may also
affect some properties of WD+MS systems, e.g. the mass transfer
rate, the critical accretion rate, the thermonuclear reaction rate
etc, leading to a different birth rate of SNe Ia.

\subsection{2002ic-like supernovae}\label{subs:6.3}
Until SN 2002ic was discovered (\citealt{HAM03}), it was long
believed that there are no hydrogen lines in the spectra of SNe
Ia. The strong hydrogen lines in the spectra of SN 2002ic were
explained by the interaction between the SN ejecta and the
circumstellar material (CSM) (\citealt{HAM03}). The CSM is
aspheric around the SN Ia from spectropolarimetry data
(\citealt{DEN04}; \citealt{WAN04}) and has a mass of
0.5-6$M_{\odot}$ (\citealt{WAN04}; \citealt{CHU04};
\citealt{UEN04}; \citealt{KOT04}). The shape of the light curve
showed that there was a delay between the explosion and the
interaction, indicating a cavity between the progenitor and the
CSM (\citealt{WOO04}; \citealt{WOO06}). Recently, two co-twins of
SN 2002ic (SN 2005gj and SN 2006gy) were also found
(\citealt{ALD06}; \citealt{OFE07}). \citet{BENETTI06} argued that
SN 2002ic is not a Type Ia supernova, but a Type Ic SN surrounded
by a structured H-rich CSM, possibly asymmetric. However, recent
spectral comparison between SN 2005gj, SN 2002ic and SNe Ia
provided further evidence of Type Ia rather than Ic for SN 2002ic
(\citealt{PRI07b}). Thus, we still take SN 2002ic as a SN Ia. To
explain these rare objects, many models were suggested and here we
just list some of them: \citealt{HAC99b}; \citealt{HAM03};
\citealt{LR03}; \citealt{CHU04}; \citealt{HAN06}.

Among all the models listed above, the model of \citet{HAN06} is
the best model to match with the observation at present,
especially about the birth rate of these rare objects and the
delay time (\citealt{ALD06}; \citealt{PRI07b}). In the scenario of
\citet{HAN06}, SN 2002ic might be from the WD + MS channel, where
the CO WD accretes material from its relatively massive companion
($\sim 3.0M_{\odot}$), and increases its mass to $\sim 1.30
M_{\odot}$ before experiencing a delayed dynamical instability.
From this scenario, we examined the initial parameters for
2002ic-like SNe Ia by assuming that the CO WD can increase its
mass to 1.378 $M_{\odot}$ and explode as a SN 2002ic-like case if
the mass of the CO WD exceeds 1.30 $M_{\odot}$ before the delayed
dynamical instability. As shown in Figs. \ref{fig1} and
\ref{fig2}\footnote{See my personal web site {\sl
http://www.ynao.ac.cn/$^{\rm \sim}$bps /download/xiangcunmeng.htm}
to see the cases with other metallicities.}, the parameter space
for the SN 2002ic-like case becomes larger with metallicity. For
$Z=0.0001$, no systems become 2002ic-like supernovae. This means
that \emph{\sl the SN 2002ic-like case could not occur in
extremely low-metallicity environments if the model in
\citet{HAN06} is appropriate}. Observationally, the host galaxy of
SN 2006gy has a solar metallicity, while there is no information
on the metallicity of SN 2002ic since any association between SN
2002ic and the galaxies near SN 2002ic have been ruled out
(\citealt{HAM03}; \citealt{KOT04}). However, we may speculate that
the metallicity of SN 2002ic is likely to be higher than $0.0001$
from the redshift of $z=0.066$. For the host galaxy of SN 2005gj,
we only know $Z/Z_{\odot}< 0.3$ (\citealt{ALD06}), which might not
be as low as $0.0001$. Our model predicts that no SN 2002ic-like
events would be observed in extremely low metallicity environments
or at very high redshift, which can be tested with an enlarged
sample of 2002ic-like supernova in the future.

\section{Summary and conclusion}\label{sect:7}
Adopting the prescription of \citet{HAC99a} for the mass accretion
of CO WDs and assuming that the prescription is valid for all
metallicities, we have studied the progenitors of SNe Ia in the
single degenerate channel (WD+MS) via detailed binary evolution
calculations. The initial parameters for SNe Ia in the ($\log
P^{\rm i}, M_{\rm 2}^{\rm i}$) plane are obtained for ten
metallicities. One can download the FORTRAN code for the contours
leading to SNe Ia at {\sl http://www.ynao.ac.cn/$^{\rm
\sim}$bps/download/xiangcunmeng.htm}.

We found that the contours for SNe Ia move to high mass and long
orbital period with increasing metallicity. The minimum initial
mass of the CO WDs for SNe Ia sharply decreases with metallicity
and the difference of the minimum initial mass between $Z=0.06$
and $Z=0.0001$ is as large as 0.24 $M_{\odot}$. Incorporating the
binary evolution calculation results in this paper into Hurley's
rapid binary evolution code, we have studied the evolution of the
birth rates of SNe Ia with time. The Galactic birth rate from the
WD+MS channel is lower than but still comparable, within a factor
of a few, to that inferred from observations. For the cases of a
single starburst, the SNe Ia explosions occur earlier and the peak
value of the birth rate is larger for a high metallicity. The
distributions of the initial masses of CO WDs, the initial masses
of secondaries, the initial periods and the initial separations
evolve significantly with metallicity. Based on the model of
\cite{HAN06}, we predict that 2002ic-like supernovae would not
occur in extremely low-metallicity environments.

\section*{Acknowledgments}

We thank Dr. Richard Pokorny for his kind help in improving the
language of this paper. This work was in part supported by Natural
Science Foundation of China under Grant Nos. 10433030, 10521001,
2007CB815406 and 10603013.


\label{lastpage}

\end{document}